\documentclass{PoS}
\usepackage{amsmath}
\usepackage[caption=false]{subfig}
\usepackage{graphicx}
\usepackage{placeins}
\usepackage{xspace}
\usepackage{floatrow}
\usepackage{wrapfig}

\title{Neutrino source searches and a realtime neutrino alert stream in the southern sky with IceCube starting tracks}

\ShortTitle{IceCube Southern Sky Starting Tracks}

\author{
The IceCube Collaboration\footnote{For collaboration list, see PoS(ICRC2019) 1177.}\\
{\itshape \href{http://icecube.wisc.edu/collaboration/authors/icrc19_icecube}{http://icecube.wisc.edu/collaboration/authors/icrc19\_icecube}}\\
E-mail: \email{sarah.mancina@icecube.wisc.edu, msilva@icecube.wisc.edu}
}

\abstract{

IceCube analyses which look for an astrophysical neutrino signal in the southern sky face a large background of atmospheric muons and neutrinos created by cosmic ray air showers. 
By selecting starting events in the southern sky, atmospheric muons and neutrinos with accompanying muons are rejected, producing a sample with high astrophysical neutrino purity at lower energies than northern sky samples. 
Our new selection method looks for muon tracks from a neutrino interaction with a vertex contained inside the detector volume by using the good pointing resolution of the track morphology to create an event specific veto region in the detector to reject entering tracks. 
This starting track event selection has a high astrophysical neutrino purity above 10 TeV at declinations less than -30$^{\circ}$ which makes it ideal for use as a southern sky realtime neutrino alert stream. 
We will discuss neutrino point source searches using this event selection and look at the advantages of the starting track alert stream for multimessenger astrophysics.\\

\vspace{4mm}
{\bfseries Corresponding authors:}
\speaker{Sarah Mancina}$^{1}$, Manuel Silva$^{1}$\\
{$^{1}$ \itshape University of Wisconsin-Madison}\\

}

\FullConference{36th International Cosmic Ray Conference -ICRC2019-\\
		July 24th - August 1st, 2019\\
		Madison, WI, U.S.A.}

\begin{document}
\raggedbottom
\section{Introduction}\label{sec:intro}
The IceCube South Pole Neutrino Observatory looks for astrophysical neutrinos that lie among a large background of muons created in cosmic ray air showers in the atmosphere. 
IceCube is an array of 86 strings--each with 60 digital optical modules (DOMs)--drilled into over a cubic kilometer of ice at the geographic south pole.
The earth can be used to reduce the muon background by looking for up-going events which pass through the interior of the earth which is nearly transparent to neutrinos. 
This technique is used in the realtime neutrino alert stream that discovered the IceCube 170922A event which was found in the direction of the flaring blazar TXS 0506+056~\cite{IceCube:2018cha}.
However, with this method of background reduction, there is still a significant atmospheric neutrino background even at energies up to 100TeV~\cite{Aartsen:2015rwa}.
To study the astrophysical neutrino sky at energies below 100TeV, it is necessary to reduce our atmospheric neutrino background even further.
By looking for neutrino events with their interaction vertex inside the detector, we have the ability to reject atmospheric neutrinos accompanied by muons from the same cosmic ray showers in the southern sky~\cite{Arguelles:2018awr}.
With this technique, we can obtain a sample of higher purity astrophysical neutrinos in the 1TeV - 100TeV neutrino energy range.

For a neutrino to be observed, it must first interact through the weak force to produce a high energy, relativistic charged particle. Different particle interactions can create different morphological shapes in the IceCube detector. 
A charged-current muon neutrino interaction will create a muon that will travel in a relatively straight path before it decays or is absorbed.
We refer to muon-like morphologies as tracks since they can traverse over a kilometer, leaving a long trace of Chernkov emission in the detector.
Charged-current electron neutrino interactions and neutral current interactions create electromagnetic and hadronic showers.
These shower morphologies appear much more isotropic in their light deposition than tracks and are referred to as cascades.

Previous IceCube starting track event selections have used the outer layers of the detector as a veto region, in which if light is deposited, the event is rejected~\cite{Aartsen:2013jdh}. 
To obtain higher purity at neutrino energies below 100TeV, the size of the veto layers can be adjusted based on the total amount of charge deposited by the event~\cite{Aartsen:2014muf}.
These veto region selections have been used to identify a diffuse astrophysical flux~\cite{Aartsen:2013jdh, Aartsen:2014muf}. 
However, they greatly reduce the fiducial volume of the detector at lower energies and create a cascade dominated sample for which the angular resolution is around 5$^{\circ}$ - 20$^{\circ}$.
The event selection technique presented here uses the better angular resolution of tracks to evaluate each event uniquely and assess if the track is starting inside the detector.

This work consists of our new starting track event selection and its application to searches for neutrino sources in the southern sky at energies of 1TeV - 200TeV. 
The following sections will first focus on the event selection technique, then discuss sensitivities of time-integrated astrophysical neutrino source searches, and finally cover the proposed near-realtime neutrino alert stream using this selection.
This event selection has interesting implications for diffuse astrophysical neutrino spectrum analyses~\cite{Silva:2019icrc_estes}; however, we will focus on the strengths of this selection for studying southern sky sources at medium neutrino energies.
\newpage
\section{Starting Track Event Selection}\label{sec:estes}
\subsection{Incoming Muon Veto}
\begin{wrapfigure}{R}{0.45\textwidth}
  \begin{center}
    \includegraphics[width=1.\textwidth]{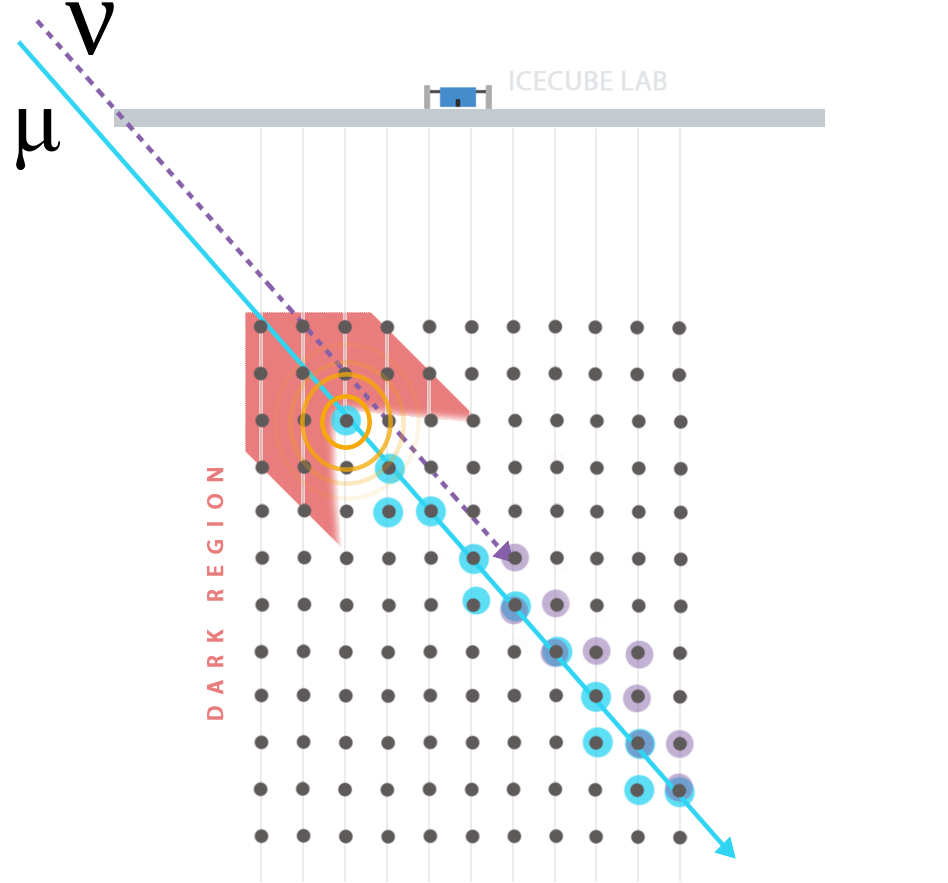}
  \end{center}
  \caption{Diagram of the dark region definition for an atmospheric neutrino being vetoed due to the incoming muon it is accompanied by.}
  \label{fig:atmo_nu_veto}
\end{wrapfigure}
To reject background atmospheric neutrino events, our incoming muon veto technique uses the expected light deposition as a function of time of minimum ionizing muons to calculate the probability that an event could have been a cosmic ray air shower muon.
An IceCube event is composed of photon hit information from the DOMs, which can be used to reconstruct the path of the particle through the ice.
Given a track reconstruction and set of DOM hits, we are able to define two regions around the track: the dark region, where there is no light consistent with the muon track timing, and the muon region, where there are hits observed consistent with the timing of the muon hypothesis. 
Then, we can use light in the muon region to scale the expected average luminosity of the track and calculate the probability, $p_{\text{miss}}$, that dark region DOMs did not observe charge if the event was an incoming muon.

\begin{figure}[tb]
    \centering
    \includegraphics[width = .6\textwidth]{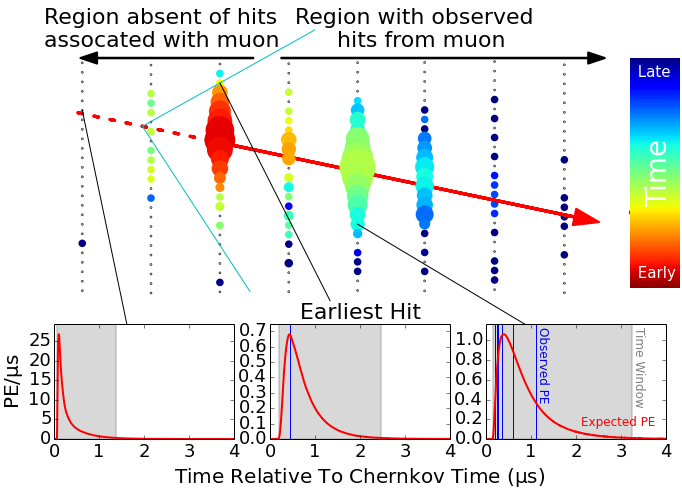}
    \caption{A two-dimensional diagram of our starting track event selection technique. Each DOM has an expected charge distribution as a function of time (red curve). For all DOMs within 350m of the track, if the observed photo-electrons (blue lines) are seen within the time window (grey box) they are classified as consistent with the muon track hypothesis.}
    \label{fig:panop}
\end{figure}

We first delineate the muon region and dark regions as illustrated in figure \ref{fig:atmo_nu_veto} and \ref{fig:panop}.
Given a reconstructed track hypothesis, we can calculate the expected photo-electron (PE) yield over time seen by DOMs from a minimum-ionizing muon. The photon expectation includes modeling of the Antarctic ice properties.
A time window is constructed for each DOM around the peak of the expected charge to contain 99\% of the total charge.
If a DOMs observed pulses during its time window, we categorize that DOM as seeing hits consistent with the track hypothesis.
The path between DOMs and the track are traced along the Cherenkov angle to find the point at which unscattered light would have been emitted.
We select the first point along the track where Cherenkov light could have been emitted and hit one of the DOMs which observed light consistent with the track hypothesis to determine the start of the track.
To define our two regions we first consider DOMs within 350m of the track. Then, we construct a Cherenkov cone for light radiating out of the reconstructed interaction vertex.
The muon region is all DOMs that within of the cone and 350m of the track, and the dark region is all DOMs that lie behind the cone and within 350m of the track.

Second, we must calculate the probability that the DOMs in the dark region did not observe charges from our track hypothesis.
For a given DOM, we expect the probability of observing a number of PE, $k$, to be poisson distributed with an mean of $\lambda$: $p(\lambda, k) = e^{-\lambda}\lambda^{k}/k!$.
We can define the probability that dark region DOMs did not see charges, $p_{\text{miss}}$, as the product of probabilities that those DOMs saw zero charge from our track hypothesis:
\begin{equation}
    p_{\text{miss}} = \prod_i^{\text{Dark Region DOMs}} \log(p(\lambda_i, k = 0))
    \label{pmiss}
\end{equation}
where $\lambda_i$ is defined as $\lambda_i(a) = a \times \lambda_{\text{muon}} + \lambda_{\text{noise}}$.
The scale factor, $a$, is calculated by maximizing the likelihood in equation \ref{STVLLH}. 
This allows us to use the hit information in the muon region to infer the average luminosity of the muon with this simplified model of the stochastic energy losses.
\begin{equation}
    \text{LLH} = \sum_i^{\text{Muon Region DOMs}} \log(p(\lambda_i(a), k_i))
    \label{STVLLH}
\end{equation}
With the scale factor, $a$, determined we can calculate $p_{\text{miss}}$ in equation \ref{pmiss}, which is used later in the event selection. 

\subsection{Event Selection}
The starting track event selection is built around the $p_{\text{miss}}$ parameter; however, we must test several track hypothesis and use a boosted decision tree to obtain the desired purity. 
The IceCube detector strings are spaced 125m apart on a triangular gird to create a hexagonal shaped detector.
The large spacing between strings allows muons to sneak past the external layers of the detector. 
Therefore, the selection must check if the tracks could have entered into the holes in the detector by testing several possible paths through the alleyways of the detector.

First, we select events which pass through simple filters that look for muon-like events or starting events and require a minimum amount of charge deposited in the detector. 
We then calculate $p_{\text{miss}}$ for an pre-run track reconstruction and reject events with a $p_{\text{miss}}$ greater than $10^{-3}$.

After the quick initial cuts, the event selection generates test paths that a muon could take through alleyways of the detector.
A point 300m down the track from the event center of charge is defined as the shifted center of charge.
We take a table of positions along the edge of the detector in between the strings of the detector and trace a path to the shifted center of charge to create a test track.
The reduced log likelihood (rLLH) of the reconstruction is calculated for all of the test tracks and select tracks with an rLLH within 2\% of the maximum rLLH of all the tracks to calculated their $p_{\text{miss}}$.
If the minimum $p_{\text{miss}}$ of all the tracks is less than $10^{-3}$ the event is kept in the next stage.

These same best rLLH test tracks are split into 131 segments, and for each segment the expected photon yield as a function time is calculated for each DOM to create the time windows used in calculating $p_{\text{miss}}$. 
By splitting the track into segments, the new timing information provides a more accurate estimation for the start of the muon region and affects $\lambda_{\text{muon}}$ of $\lambda_i$ in equation \ref{pmiss}.
A cut is made on the minimum $p_{\text{miss}}$ obtained in this step requiring it to be less than $10^{-5}$.

The final test of alleyways is a finer scan of track directions around the top rLLH test tracks from the coarser scan. 
The shifted center of charge, zenith, and azimuth of the top test tracks are slightly shifted around to test a total of 1625 fine search tracks around each initial test track from the coarser search.
Again, tracks with an rLLH within 2\% of the max rLLH are kept and their $p_{\text{miss}}$ calculated. 
The track is kept if, after splitting the tracks into 131 segments like above, the minimum $p_{\text{miss}}$ is less that $10^{-5}$.

The initial track reconstruction is then fed to a stochastic muon energy loss reconstruction. For events with a zenith of less than 80$^{\circ}$ ($\delta < -20^{\circ}$), the energy loss information, $p_{\text{miss}}$ values, and other quality information are fed to a boosted decision tree trained to distinguish between muon and starting neutrino events~\cite{Silva:2019icrc_estes}.
The parameters that are most important in the BDT are the fraction of energy lost in the first reconstructed loss and the distance of the reconstructed start of the track to the edge of the detector.
Some simple quality cuts are run on the up-going events.
This brings the event selection to its final level.

\begin{figure}
    \centering
    \includegraphics[width = .49\textwidth]{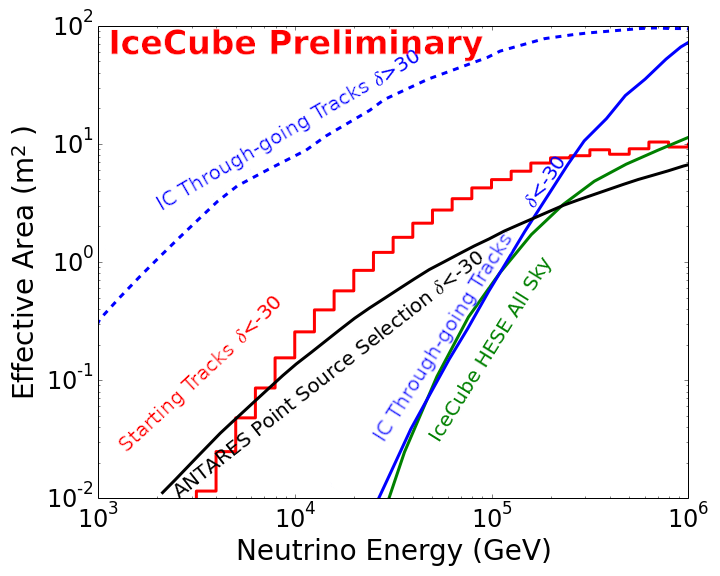}
    \includegraphics[width = .49\textwidth]{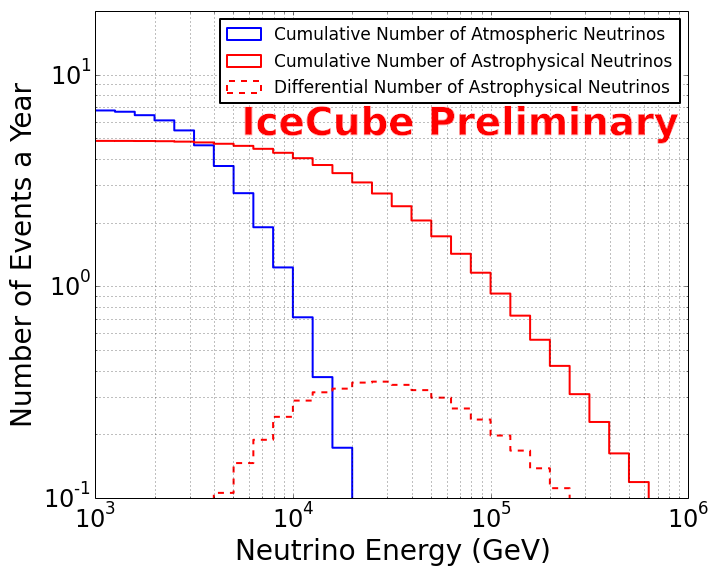}
    \caption{Left: Effective area of the starting track event selection in the southern sky (red) binned in true neutrino energy. The effective area is greater than the IceCube through-going event selection (blue)~\cite{Aartsen:2016oji} in the southern sky at energies below 200TeV. Right: Cummulative and differential distribution of events per year as a function of true neutrino energy from $-90^{\circ}$ to $-30^{\circ}$ assuming the flux from~\cite{Aartsen:2014muf}. The atmospheric self-veto is responsible for the suppression of atmospheric neutrinos at TeV scales.}
    \label{fig:eff_a}
\end{figure}

At the final level of our selection we are left with a relatively pure sample of astrophysical neutrinos in the southern sky. We expect less than one atmospheric muon per year in the event selection as seen in Table \ref{tab:eventrate}.
\begin{table}
\centering
\begin{tabular}{c|c c c}
     & Atmospheric $\mu$ & Atmospheric $\nu$ & Astrophysical $\nu$ \\
     \hline
    Up-going ($\delta > -20^{\circ}$) & 0 & 127 & 8 \\
    Down-going ($\delta \leq -20^{\circ}$) & 0.8 & 33 & 8 
\end{tabular}
\caption{Final level expected events per year assuming an astrophysical flux of $2.06 \times 10^{-18} \left(\frac{E}{100TeV} \right)^{-2.46}$ [GeV$^{-1}$ cm$^{-2}$ s$^{-1}$ sr$^{-1}$] from~\cite{Aartsen:2014muf}.
The atmospheric neutrino self-veto effect in the southern sky is modeled by forcing neutrinos in CORSIKA simulation to interact~\cite{Heck:1998vt}.}
\label{tab:eventrate}
\end{table}
This event selection has the largest neutrino effective area at declinations of less that 30$^{\circ}$ between 8TeV and 200TeV (Figure~\ref{fig:eff_a}).
Other IceCube event selections are forced to cut strongly on energy in the southern sky in order to reduce the muon background~\cite{Aartsen:2016oji}.
The northern sky starting track event selection has a similar effective area as in the southern sky, which is not competitive with the regular through-going point source tracks event selection and there is a large overlap of events in the two selections in the northern sky.
The starting track event selection also suppresses atmospheric neutrinos in the 10TeV - 100TeV energy range. This results in a higher purity of astrophysical neutrinos at these energies.

The direction and energy reconstruction resolution are affected by the nature of starting tracks. The average angular error is 1.7$^{\circ}$. 
The angular error is highly correlated with the length of the track in the detector, and southern events ($\delta$ < -70$^{\circ}$) tend to have shorter lengths due to stricter requirements on the distance of the start of the track to the edge of the detector from the BDT.
Therfore, the angular resolution is slightly worse in these most southern declinations with a median angular error of 3.0$^{\circ}$.
The neutrino energy reconstruction resolution is 0.25 in the $\log_{10}(E_{\nu})$ space at all neutrino energies from 1TeV to 1PeV. The through-going track selection uses a reconstruction that has a muon energy resolution of 0.22 in the $\log(E_{\mu})$ space~\cite{Abbasi:2012wht}. The improved neutrino energy resolution for starting tracks is due to the addition of information from the hadronic cascade at the start of events.

\section{Neutrino Source Searches}\label{sec:nusource}
\begin{wrapfigure}{R}{0.45\textwidth}
  \begin{center}
    \includegraphics[width=1.\textwidth]{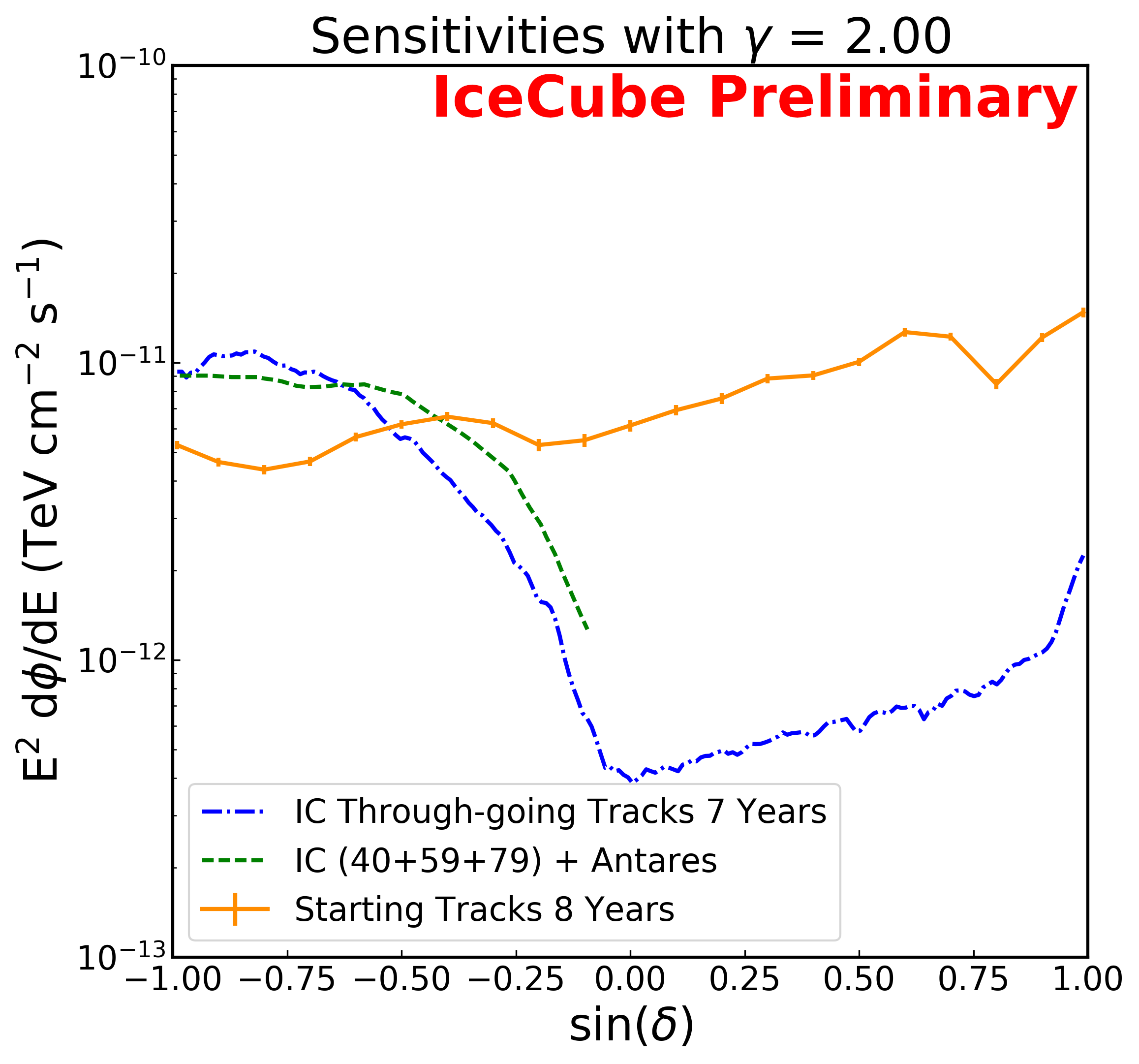}
  \end{center}
  \caption{Pre-trial sensitivities for the starting tracks compared to the sensitivities from~\cite{Aartsen:2016oji} and~\cite{Adrian-Martinez:2015ver}.}
  \label{fig:sens_g2.0}
\end{wrapfigure}
For our time-integrated neutrino source searches, we plan to do an all-sky search, catalog search, stacking search, and galactic plane template fit. To search for point-like sources we use a unbinned maximum likelihood method to estimate the number of signal neutrinos at the potential source location in the sky like previous neutrino searches~\cite{Aartsen:2016oji}. 

In the all-sky search, we scan a HEALPix grid of pixels that are $1.6 \times 10^{-5}$  steradians in area. To calculate the pre-trial sensitivities, first, we create a background test statistic distribution and find the median test statistics for each test declination. Then, we inject signal events with a power law energy distribution and find the flux normalization for which 90\% of the injected trials are larger than the median test statistic. 

In figure \ref{fig:sens_g2.0} the sensitivities are shown for a source with a spectral index of 2 for the normalization at 100TeV. 
As expected, the starting track sensitivity becomes competitive in the southern sky where the through-going tracks become less sensitive.
The improved sensitivity is more apparent when the source spectral index increases to 2.5 or 3 because we then expect more neutrinos at energies below 100TeV and the starting track event selection's effective area is better than the through-going tracks sample in the mid-TeV range (figure \ref{fig:eff_a}).
Similarly the event selection also improves current limits in the southern sky when an exponential energy cutoff is applied to the source energy spectrum.
There are plans to combine both event selections to increase sensitivity in the future.

For the galactic template analysis we use the unbinned likelihood with signal subtraction as used in \cite{Aartsen:2017ujz}. 
In this method we test for a signal from two models of the diffuse galactic plane neutrino emission: the Fermi $\pi^0$ model~\cite{Ackermann:2012pya} and the KRA$\gamma$ model~\cite{Gaggero:2015xza}. 
A comparison to the previously published results with the 7 year template fit is shown in table \ref{tab:templates}. The starting tracks have a greater sensitivity due to the location of denser regions of the galaxy in the southern sky and the shape of the energy spectrum of the galactic plane emission models.
\begin{table}
    \centering
    \begin{tabular}{c|ccc}
         &  Fermi $\pi^0$~\cite{Ackermann:2012pya} & KRA$\gamma$~\cite{Gaggero:2015xza} & KRA$\gamma$~\cite{Gaggero:2015xza} \\
         & $(\text{TeV}^{-1} \text{cm}^{-2} \text{s}^{-1})$ & (5 PeV Cutoff) & (50 PeV Cutoff)\\
         \hline
         Starting Tracks (8 Years) & $2.45 \times 10^{-11}$ & $45.9\% \times \text{KRA}\gamma$ & $35.1\% \times \text{KRA}\gamma$ \\
         Through-Going Tracks (7 Years)~\cite{Aartsen:2017ujz} & $2.97 \times 10^{-11}$& - & $79\% \times \text{KRA}\gamma$
    \end{tabular}
    \caption{Comparison of the galactic plane template model sensitivities for this starting track event selection and the published icecube through-going tracks event selection. For the Fermi $\pi^0$ template, a power law with a spectral index of 2.5 is assumed and for the KRA$\gamma$ template the spectrum shown in~\cite{Gaggero:2015xza} is used.}
    \label{tab:templates}
\end{table}

\section{Realtime Alert Stream}\label{sec:realtime}
Running the full selection at the South Pole in realtime is not possible due to its time and memory requirements, so a modified version is run instead.
The modified selection only looks for neutrinos from the southern sky ($\delta \leq -20^{\circ}$). 
If an event passes the modified version at the south pole, the candidate event is then sent North to be processed by the full selection.
The modified selection requires a larger amount of energy deposited in the detector and a longer track length than the offline version; therefore, events that pass both the modified and full selection have a higher purity and quality than the regular full offline sample.

The modified online selection triggers on average 16.8 times per day. Most of these events are expected to be muons. 
From simulation, we estimate approximately 17.9 atmospheric neutrinos per year and 5.5 astrophysical neutrinos per year with 50\% signalness or greater to pass through the modified selection assuming the astrophysical flux from~\cite{Aartsen:2014muf}. 
The signalness is calculated by taking the percentage of astrophysical neutrinos to the total number of neutrinos from simulation that lie in the same reconstructed energy and declination at the final event selection level.
The 50\% signalness events have energies in the 10TeV to 200TeV range, which is a lower range than the currently running alerts~\cite{Aartsen:2016lmt}.
Due to the low energy range, these events could be of great interest for galactic transient events, especially since this stream is looking at the southern sky where some of the most active parts of the galaxy are located.

\section{Concluding Remarks}
The starting track event selection takes advantage of the pointing resolution of IceCube muon tracks to evaluate event by event the probability that the track is an incoming cosmic ray muon.
This allows us to reject not only muons from cosmic rays, but also atmospheric neutrinos with accompanying muons.
Our new selection can probe the southern sky at lower neutrino energies due to the supression of atmospheric neutrinos.
The high purity of our event selection makes it a great candidate for a realtime neutrino alert stream.
The realtime starting track events are anticipated to be added to IceCube's community alerts soon.

\bibliographystyle{ICRC}
\bibliography{references}

\end{document}